\documentclass[aps,prb,reprint,superscriptaddress]{revtex4-2}
\usepackage{CJK}
\usepackage{gensymb}
\usepackage{graphicx}
\usepackage{amssymb,amsmath,mathptmx}
\usepackage[colorlinks=true,citecolor=blue,linkcolor=blue,urlcolor=blue,bookmarks=false]{hyperref}

\usepackage{tikz}
\usetikzlibrary{arrows,snakes,positioning,shapes,patterns}

\def\unit#1{\mathord{\thinspace\rm #1}}

\newcommand{\sgn}{\text{sgn}}

\begin{document}

\begin{CJK*}{UTF8}{}
\title{Manipulating electron waves in graphene using carbon nanotube gating}

\author{Shiang-Bin Chiu (\CJKfamily{bsmi}{邱翔彬})}
\affiliation{Department of Physics, National Cheng Kung University, Tainan 70101, Taiwan}
\affiliation{Department of Physics, National Taiwan University, Taipei 10617, Taiwan}

\author{Alina Mre\'nca-Kolasi\'nska (\CJKfamily{bsmi}{柯琳娜})}
\email{alina.mrenca@fis.agh.edu.pl}
\affiliation{Department of Physics, National Cheng Kung University, Tainan 70101, Taiwan}

\affiliation{AGH University of Science and Technology, Faculty of Physics and Applied Computer Science, al. Mickiewicza 30, 30-059 Krak\'ow, Poland}

\author{Ka Long Lei (\CJKfamily{bsmi}{李嘉朗})}
\affiliation{Department of Physics, National Cheng Kung University, Tainan 70101, Taiwan}

\author{Ching-Hung Chiu (\CJKfamily{bsmi}{邱靖鈜})}
\affiliation{Department of Physics, National Cheng Kung University, Tainan 70101, Taiwan}

\author{Wun-Hao Kang (\CJKfamily{bsmi}{康文豪})}
\affiliation{Department of Physics, National Cheng Kung University, Tainan 70101, Taiwan}

\author{Szu-Chao Chen (\CJKfamily{bsmi}{陳思超})}
\affiliation{Department of Physics, National Cheng Kung University, Tainan 70101, Taiwan}

\author{Ming-Hao Liu (\CJKfamily{bsmi}{劉明豪})}
\email{minghao.liu@phys.ncku.edu.tw}
\affiliation{Department of Physics, National Cheng Kung University, Tainan 70101, Taiwan}

\date{\today}

\begin{abstract}
Graphene with its dispersion relation resembling that of photons offers ample opportunities for applications in electron optics. 
The spacial variation of carrier density by external gates can be used to create electron waveguides, in analogy to optical fiber, with additional confinement of the carriers in bipolar junctions leading to the formation of few transverse guiding modes.
We show that waveguides created by gating graphene with carbon nanotubes (CNTs)
allow obtaining sharp conductance plateaus,
and propose applications in the Aharonov-Bohm and two-path interferometers, and a pointlike source for injection of carriers in graphene. Other applications can be extended to Bernal-stacked or twisted bilayer graphene or two-dimensional electron gas.
Thanks to their versatility, CNT-induced waveguides open various possibilities for electron manipulation in graphene-based devices.
\end{abstract}

\maketitle
\end{CJK*}

\section{Introduction}
Graphene's linear dispersion relation, resembling the one of photons, inspired plethora of applications of graphene for electron optics. External gates can be used to locally tune the Fermi energy
, which, by analogy to optics, plays the role of the refractive index. Moreover, graphene can be smoothly modulated between electron and hole conduction, thus it is possible to create junctions between regions of opposite polarity. 
Thanks to this flexible control of the carrier density, electrostatically defined optical elements such as lenses \cite{Cheianov2007veselago, Lee2015, Liu2017beams, Boggild2017, Brun2019}, collimators \cite{Cheianov2006, Wang2019}, Fabry-P\'erot  \cite{Young2009fp, Rickhaus2013fp, Grushina2013fp} and Mach-Zehnder interferometers \cite{Wei2017, Jo2021} or microcavities \cite{Bardarson2009, Wurm2011, Zhao2015, Schrepfer2021dirac, Brun2022} are realizable in graphene and have been widely explored both theoretically and experimentally. 
Furthermore, unlike photons, carriers in graphene are charged, which opens up opportunities for applications beyond the regular optics, including manipulation with external magnetic field for transverse magnetic focusing \cite{Taychatanapat2013, Chen2016electron_optics, Berdyugin2020} or the Aharonov-Bohm effect \cite{Russo2008, Deprez2021, Ronen2021}. 

\begin{figure}[b]
\includegraphics[width=\columnwidth]{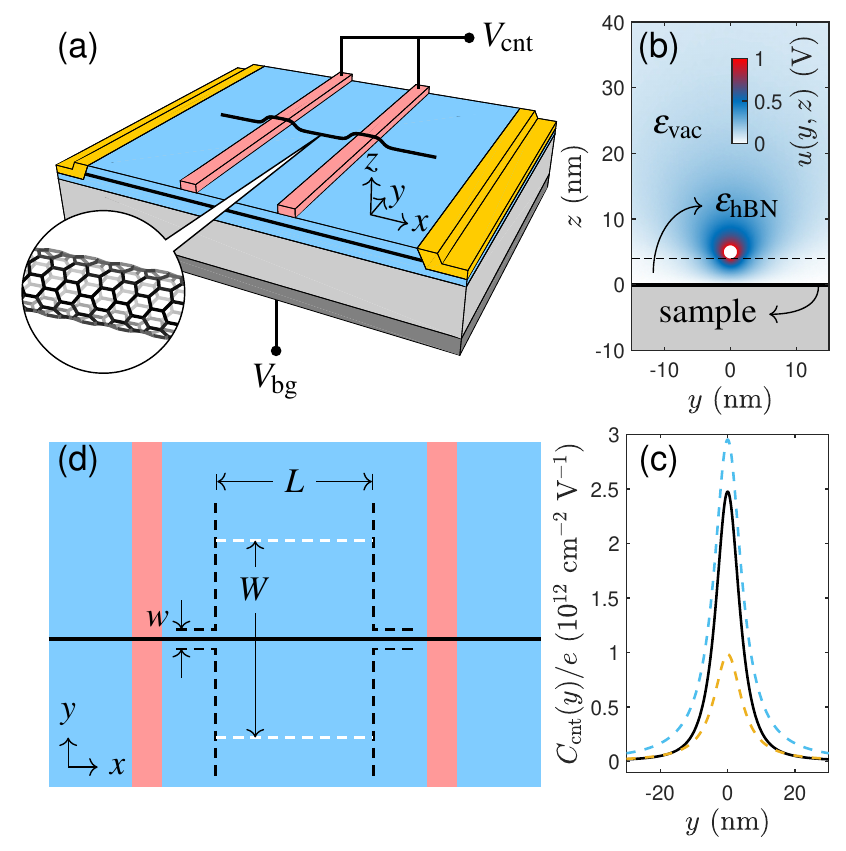} 
\caption{(a) Sketch of the considered electron system gated by CNT.
(b) Spatial profile of the electrostatically simulated electric potential $u(y,z)$ considering a grounded sample subject to a CNT gate of radius $r_\mathrm{cnt}=1\unit{nm}$ applied with $V_\mathrm{cnt}=1\unit{V}$. (c) Capacitance profile (solid line) corresponding to (b), where dielectric constant $\epsilon_\mathrm{hBN}=3$ below the nanotube and $\epsilon_\mathrm{vac}=1$ elsewhere are considered. The blue (orange) dashed line shows the analytical result for uniform dielectric constant $\varepsilon_\mathrm{hBN}=3$ ($\varepsilon_\mathrm{vac}=1$). (d) Top view of the device. The dashed lines show the area considered in the transport calculation. } 
\label{fig:fig1}
\end{figure}

The possibility of spatial variation of the potential profile 
can be utilized to form electron waveguides, with three regions of varying carrier density being counterparts of materials with different refractive indices in the optical fiber. By analogy to the total internal reflection of light in the waveguide core having the refractive index higher than the surrounding cladding, in a channel induced electrostatically, electrons incident below the critical angle are trapped and propagate along the channel \cite{Williams2011}. In addition to this optical fiber guiding (OFG), when the polarity at the interface is inverted, a bipolar pnp or npn junction is formed which can impose additional carrier confinement, leading to formation of few guiding modes.

Few-mode guiding in graphene has been widely discussed in theory \cite{Pereira2006g, Beenakker2009g, Zhang2009g, Hartmann2010smooth, Hartmann2014g, Liu2015scalable, Shah2019g} and successfully realized in experiments \cite{Rickhaus2015guiding, Kim2016valley} which employed narrow electrostatic gates. 
However, in waveguides induced by electrostatic gates the interface is bound to be irregular.
The possibility to circumvent these limitations is by using a carbon nanotube (CNT) as a gate, which can induce a sharp and regular interface. Moreover, the CNT shape can be controlled to some extent \cite{Wang2006, Joselevich2009} allowing for flexible design of the waveguide geometry. Recent advancement in the fabrication of nanostructures, and, in particular, efficient transfer and manipulation of CNTs for the assembly of nanodevices \cite{Huang2005, Marganska2019, Otsuka2021, Ozdemir2021, Guo2022}, opens up possibilities for precise control over the CNT position and orientation.
Using CNT as a gate for graphene has been proposed in theoretical works \cite{Lyo2003coulombdrag, Hartmann2020cnt2, Hartmann2020cnt3} as well as realized experimentally in the capacitive measurement of graphene's local density of states \cite{Cheng2019cntguiding} and Coulomb drag between graphene and CNT \cite{Anderson2021}. However, no transport investigations of guiding by a CNT-induced channel have been conducted so far. 

In this work, we consider the quantum transport of carriers in a system gated by charged CNT [Fig.\ \ref{fig:fig1}(a)]. We demonstrate the versatility of CNT-induced guiding channels in graphene, which can be utilized to form extremely narrow and sharp channels, electrostatically defined quantum rings \cite{Kim2016valley, Wei2018gab}, pointlike sources \cite{Liu2017beams}, interferometers, and other building blocks for nanodevices. 
Their application is not limited to single-layer graphene (SLG), and, as we show in the following, it can also be utilized in other materials, including Bernal-stacked bilayer graphene (BLG), decoupled twisted bilayer graphene (dtBLG), and semiconductor nanostructures hosting two-dimensional electron gas (2DEG).

\section{Quantized electron waveguide}

\subsection{Electrostatics }
In the following, we study two-dimensional systems (SLG, BLG, dtBLG and 2DEG) placed above a global back gate at voltage $V_\mathrm{bg}$ 
and gated from the top by a CNT at voltage $V_\mathrm{cnt}$. 
Figure \ref{fig:fig1}(a) shows the 3D design of the considered device. 
Although the experimental design differs between graphene devices and 2DEG, here for the sake of comparison we consider the same device geometry for each system: a graphene system sandwiched between two hBN layers [blue in Fig.\ \ref{fig:fig1}(a)] and placed on a SiO$_2$ substrate [light gray in Fig.\ \ref{fig:fig1}(a)], or 2DEG embedded in a medium with equivalent dielectric constants as those in the graphene device. The CNT is connected to electrodes marked in pink, and graphene to metallic contacts marked in yellow.
The CNT is separated from graphene by an hBN sheet $d_{t}=4$ nm thick, with the dielectric constant $\varepsilon_\mathrm{hBN}=3$. 
The bottom hBN layer is $d_{b}=20$ nm thick, and the SiO$_2$ substrate $d_\mathrm{SiO_2}=285$ nm thick, 
and we adopt the dielectric constant for SiO$_2$ $\varepsilon_\mathrm{SiO_2}=3.8$.
The back gate capacitance is obtained from the parallel-plate capacitor model, $C_\mathrm{bg}/e = \varepsilon_0/e \left(  d_{b}/\varepsilon_\mathrm{hBN} +  d_\mathrm{SiO_2}/\varepsilon_\mathrm{SiO_2} \right)= 6.7676\times 10^{10}\unit{cm^{-2} V^{-1}}$, where $\varepsilon_0$ is the vacuum permittivity, and $-e$ is the electron charge.

For the electrostatic modeling of a straight CNT placed along the $x$ direction, 
 instead of a full three-dimensional structure we assume the system is invariant in the $x$ direction, and model the potential profile in the transverse direction only, performing 2D electrostatic simulation in the $y$-$z$ coordinates. 
The electric potential distribution $u(y, z)$ induced by the charged CNT  for $V_\mathrm{cnt}=1$ V is shown in Fig.~\ref{fig:fig1}(b).
The numerically obtained $C_\mathrm{cnt}(y)/e$ is presented in Fig.~\ref{fig:fig1}(c) \cite{Liu2013multigated}. For comparison, the orange (blue) dashed line shows the analytical result for a uniform dielectric constant $\varepsilon_r=3$ ($\varepsilon_r=1$), given by $C_\mathrm{cnt}(y)/e = \epsilon_0\epsilon_r 4 a_t / (y^2+a_t^2) e\log(\kappa)$, with $a_t=\sqrt{h_t^2-r_\mathrm{cnt}^2}$, $h_t=d_t+r_\mathrm{cnt}$, $\kappa=(h_t+a_t)^2/r_\mathrm{cnt}^2$. 

In the case of a curved CNT, considered in Sec.\ III, the potential profile is calculated as described in Appendix \ref{app:C1Dto2D}. 
The potential profile induced by two crossed CNTs, mentioned in Sec.\ III B, is adopted from 3D finite-element modeling with the full $x$, $y$, and $z$ dependence, which yields  $C_\mathrm{cnt}(x, y)/e$. 

Given the gates capacitance, the carrier density is calculated from
\begin{equation}
\label{eq:SLGn}
n= (C_\mathrm{bg}V_\mathrm{bg}+ C_\mathrm{cnt}V_\mathrm{cnt})/e
\end{equation}
for graphene free of intrinsic doping, where $C_\mathrm{cnt}$ is a function of coordinates, yielding a position-dependent $n$. 
We assume graphene is described by the dispersion relation $E=\pm \hbar v_F k$, where $\hbar$ is the reduced Planck constant, $v_F\approx 10^6\unit{m/ s}$ is the Fermi velocity of graphene, and we adopt $\hbar v_F \approx 3\sqrt{3}/8\unit{eV nm}$.
The on-site energy which we input into the Hamiltonian 
is calculated from 
\begin{equation}
\label{eq:SLGE}
U=-\sgn(n) \hbar v_F \sqrt{\pi |n|}.
\end{equation}

The 2DEG band structure differs from that of graphene: $E=\hbar^2 k^2/2m^*$ in the effective mass approximation,  where we use  the effective mass for GaAs $m^*=0.067 m_0$ with $m_0$ being the electron mass. The on-site energies are $U=-\pi\hbar^2n/m^*$, with $n$ given by Eq.\ (\ref{eq:SLGn}).
The BLG density and on-site energies calculation follows Ref.\ \onlinecite{Varlet2014}, and for dtBLG we adopt the self-consistent model for zero magnetic field described in Ref.\ \onlinecite{Mrenca2021}.

\subsection{Transport calculation}

Figure \ref{fig:fig1}(d) shows the device top view. For the transport calculation, to focus on the guiding effect of the CNT gate, we consider an idealized four-terminal device marked by the black dashed lines in Fig.~\ref{fig:fig1}(d). The contacts are simulated by semi-infinite leads, and the computational box is limited to a rectangle of size $L\times W = 360$ nm $\times 360$ nm, unless stated otherwise. Electrons are guided between the left source lead and the right collector lead of width $w=34$ nm. The current leaking out of the guiding channel flows to the top and bottom leads. 

The calculations are based on the tight-binding Hamiltonian
\begin{equation}
\label{eq:Htb}
H = -\sum\limits_{\left\langle {i,j} \right\rangle } \left( t_{ij} c_i^\dagger  c_j + H.c. \right) +\sum\limits_j {U({\mathbf{r}}_j)} c_j^\dagger c_j,
\end{equation}
where the operator $c_i$ ($c_i^\dag$) annihilates (creates) an electron on the $i$th  site located at $\mathbf{r}_i=(x_i,y_i)$, and the second sum contains on-site energies. 
In SLG and dtBLG, the hopping parameters $t_{ij}$ describe the nearest-neighbor hopping with $t_{ij}=t_0=3$ eV, whereas in BLG additionally the interlayer hoppings $t_{ij}=0.39$ eV between the dimer sites are included. For 2DEG $t_{ij} = \hbar^2/2m^*\Delta x^2$ where $\Delta x=1$ nm is the grid spacing, and the on-site energies contain an additional term $4 \hbar^2/2m^*\Delta x^2$. 
To model the external magnetic field $\mathbf{B}=(0,0,B)$, the hopping parameter is modified to contain the Peierls phase $t_{ij} \rightarrow t_{ij} e^{i\phi}$, with $\phi = -\tfrac{e}{\hbar} \int_{\textbf{r}_i}^{\textbf{r}_j} \textbf{A}\cdot d\textbf{r}$, where $\textbf{A}$ is the vector potential  such that $\nabla\times \mathbf{A} = \mathbf{B}$, and the integration is from the site at $\textbf{r}_i$ to the site at $\textbf{r}_j$.
To simulate real graphene devices we adopt the scalable tight-binding model \cite{Liu2015scalable}, with the scaled hopping parameter $t=t_0/s_F$ and lattice spacing $a=a_0s_F$, where $s_F$ is the scaling factor, and we use $a_0 = 1/4\sqrt{3}$ nm. 
The Hamiltonian (\ref{eq:Htb}) is applied for transport simulation within the real-space Green's function method \cite{Datta1995}, wave-function matching \cite{Kol2016transport} or using the Kwant package
\cite{Groth2014} for SLG/2DEG, dtBLG, and BLG, respectively. 
The transport energy is chosen at $E = 0$.
At zero temperature the conductance from lead $i$ to lead $j$ is calculated using the Landauer formula $G_{ji}=2e^2T_{ji}/h $, where $T=\sum_m T_{ji}^{(m)}$ is summed over the propagating modes.

\subsection{Single-layer graphene}

The on-site energy which we input into the SLG Hamiltonian 
is given by Eq.\ (\ref{eq:SLGE}). 
In single-layer graphene, it also plays the role of the refractive index, and, in analogy to optics, the refraction at the interface is described by the Snell's law $E_\mathrm{in} \sin(\theta_\mathrm{in}) = E_\mathrm{out} \sin(\theta_\mathrm{out})$, where $E_\mathrm{in}$ ($E_\mathrm{out}$) is the energy within (outside) the channel. The total internal reflection occurs when the incidence angle satisfies $\theta>\theta_c$, with $\theta_c = \arcsin(E_\mathrm{out}/E_\mathrm{in})$. Thus the OFG in the channel is possible when $|E_\mathrm{in}|>|E_\mathrm{out}|$. It is equivalent to the requirement $|k_\mathrm{in}|>|k_\mathrm{out}|$ written in terms of the wave vector within (outside) the channel $k_\mathrm{in}$ $(k_\mathrm{out})$. 
The confinement in the bipolar junction, which appears to be stronger than in OFG, is realized when $E_\mathrm{in}E_\mathrm{out}<0$. In terms of the carrier densities $n_\mathrm{in}$ and $n_\mathrm{out}$, it is equivalent to $n_\mathrm{in}n_\mathrm{out}<0$, which in our system roughly corresponds to $V_\mathrm{bg}V_\mathrm{cnt}<0$, as estimated by $n_\mathrm{in}n_\mathrm{out} = C_\mathrm{bg}V_\mathrm{bg}(C_\mathrm{bg}V_\mathrm{bg}+C_\mathrm{cnt}(0)V_\mathrm{cnt})\approx C_\mathrm{bg}V_\mathrm{bg}C_\mathrm{cnt}(0)V_\mathrm{cnt}$ since $C_\mathrm{bg}$ is two orders of magnitude smaller than $C_\mathrm{cnt}$ at its peak. 

\begin{figure}[t]
\includegraphics[width=\columnwidth]{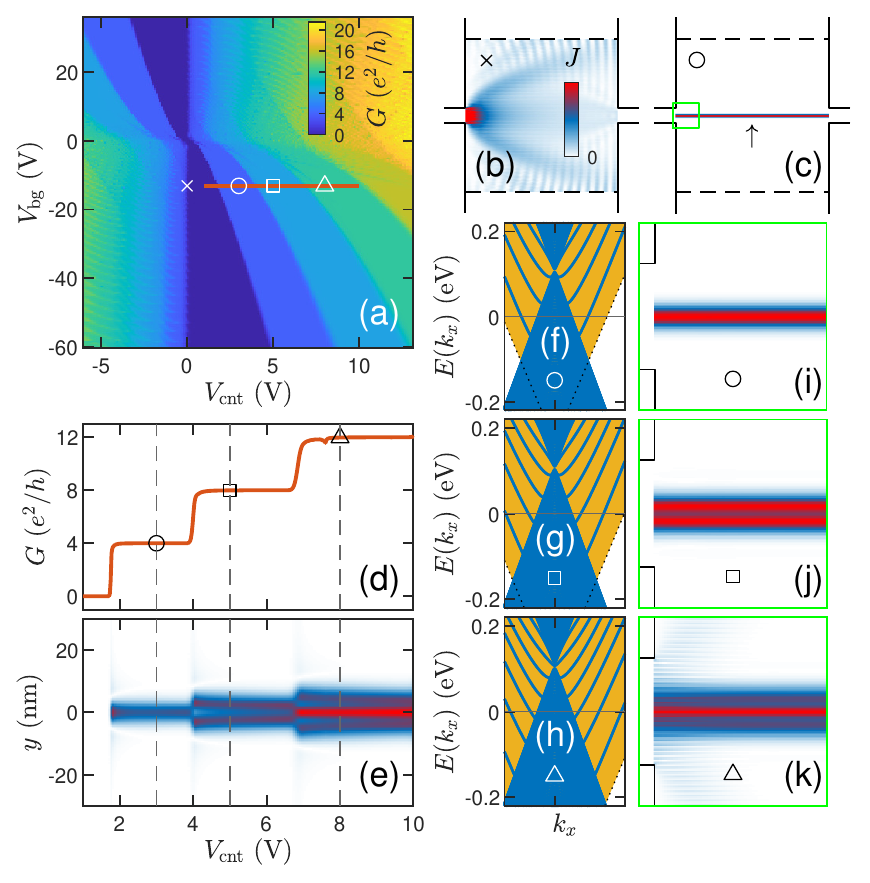} 
\caption{(a) Conductance between thin leads of SLG as a function of $V_\mathrm{cnt}$ and $V_\mathrm{bg}$,  with $V_\mathrm{bg} = -13.16\unit{V}$ marked for the rest of the panels. Spatial profile of the current density distribution (b) without the guiding channel [$V_\mathrm{cnt}=0$ marked by $\times$ on panel (a)] and (c) with the guiding channel at the lowest mode [$V_\mathrm{cnt}=3\unit{V}$ marked by $\circ$ on panel (a)]. (d) The line cut corresponding to the red line marked on (a). (e) Current density cross section at $x$ marked by the arrow in (c) as a function of $V_\mathrm{cnt}$ and $y$, illustrating the formation of quantized guiding modes. (f)--(h) Band structures and (i)--(k) current density maps showing the range marked by the green box in (c) at gate voltage points marked in (a) and (d). The yellow area in (f)--(h) indicate the region where the states confined within the channel exist. } 
\label{fig:fig2}
\end{figure}

Figure \ref{fig:fig2}(a) shows the two-terminal conductance in SLG between the narrow left and right terminals as a function of the backgate and CNT voltage, calculated with $s_F=2$.
For $V_\mathrm{bg}V_\mathrm{cnt}>0$ the junction induced by the CNT is unipolar, the confinement within the channel is relatively weak since it is only due to OFG, and the bulk states have a significant contribution to conductance. In this case conductance quantization is hardly seen, one can also spot fine oscillations which correspond to resonant states in the cavity between the vertical edges of the flake.
Figure \ref{fig:fig2}(b) presents the spatially resolved current density $J(x,y)$ for the voltages marked with a cross in Fig.\  \ref{fig:fig2}(a). Here $V_\mathrm{cnt}=0$, hence the potential profile in the device is uniform, and neither bipolar junction nor optical guiding occurs. Thus, a small part of the injected current is transmitted towards the right lead, however, a significant part flows out of the channel and escapes through the top and bottom leads.

In the quadrants $V_\mathrm{bg}V_\mathrm{cnt}<0$, bipolar junctions are formed, and the conductance shows clear plateaus as few-mode guiding is realized in the channel. In this regime, the current shows an entirely different behavior. A representative current density with a single mode available in the channel is shown in Fig.~\ref{fig:fig2}(c), corresponding to the gate voltage marked with a circle in Fig.~\ref{fig:fig2}(a), and demonstrating perfect guiding between the left and right lead. 
The conductance cross section in Fig.~\ref{fig:fig2}(d) for a fixed $V_\mathrm{bg}=-18$ V [red line in Fig.~\ref{fig:fig2}(a)] shows nearly ideal conductance quantization at values $\tfrac{4e^2}{h} M$, where $M$ is an integer, and the spacing by four arises from the spin and valley degeneracy. Figure \ref{fig:fig2}(e) shows the current density cross section at $x=0$, marked by an arrow in Fig.~\ref{fig:fig2}(c), as a function of $V_\mathrm{cnt}$. Between $V_\mathrm{cnt}\approx 2$ V and 4 V, the $M=1$ modes contribute to current. At the transition from the plateau $M=1$ to $M=2$ [Fig.~\ref{fig:fig2}(d)], the second branch of guiding modes becomes available for transport. The $M=2$ transverse modes wave functions exhibit one node in the center. In the corresponding current density in Fig.~\ref{fig:fig2}(e) two maximums can be resolved, as the current propagates in both first and second branch. 
Similarly, at the transition from $M=2$ to $M=3$, the third branch opens, and three maximums of the current density profile are resolved. In this current density map, the current is carried nearly entirely within the guiding channel, with only a small fraction flowing in the bulk when the successive guiding modes open for transport (close to the transition $M-1\rightarrow M$).

Figures \ref{fig:fig2}(f)--\ref{fig:fig2}(h) show band structures calculated for a translationally invariant ribbon of width 300 nm, obtained for the CNT voltage values marked by the corresponding symbols in Figs.~\ref{fig:fig2}(a) and \ref{fig:fig2}(d). The band structure of graphene gated by a CNT consists of Dirac cones typical for pure graphene, corresponding to the bulk graphene beyond the channel with an almost flat potential profile, and additional discrete branches arising from the confinement within the CNT-induced channel.
The energies of the states bound within the channel are within the area marked with yellow in Fig.~\ref{fig:fig2}(f)--\ref{fig:fig2}(h), delimited by $E = \pm \hbar v_F |k_x| + V_{\mathrm{out}}$ and $E = \hbar v_F |k_x| + V_{\mathrm{in}}$ [marked by the dotted lines in Fig.~\ref{fig:fig2}(f)--\ref{fig:fig2}(h)]\cite{Pereira2006g}. 
New guiding modes open at the energies for which the branches touch the Dirac cone.  Figures ~\ref{fig:fig2}(i)--\ref{fig:fig2}(k) show the representative current density maps for the cases of $M=1$, 2, and 3, corresponding to the points marked with the symbols in Figs.~\ref{fig:fig2}(a) and \ref{fig:fig2}(d). 

Figure \ref{fig:fig3}(a) shows the SLG conductance map calculated with $s_F=6$, and Fig.\ \ref{fig:fig3}(e) its cross section at $V_\mathrm{bg}=-18$ V.
 To check the applicability of higher scaling factors, we compare the cross sections of the conductance calculated with $s_F=2$ and $s_F=6$. Both lines are in a good agreement with the step only slightly shifted in $V_\mathrm{bg}$, thus we conclude with high scaling factor the results remain valid. 
 
To check the impact of disorder, we performed the calculations with the disorder potential present. These results are summarized in Appendix \ref{app:disorder}.
 
 It is worth noting that for very strong potential energy variation in space, the intervalley scattering becomes relevant in a channel along the zigzag direction
, and gives rise to intermediate plateaus $G=4e^2(M-1/2)/h$. We elaborate on this in Appendix \ref{app:plato}.

The calculations presented in this section are obtained for device with the hBN thickness of 4 nm based on the experiment \cite{Cheng2019cntguiding}. However, for a few-nanometer thin hBN and high CNT voltage, there is a risk of a dielectric breakdown \cite{Ranjan2021}. To consider a safer design, in Appendix \ref{app:thick} we present the calculations for the case of 10, 15, and 20 nm thick hBN between the SLG and the CNT. We also consider wider injection leads in Appendix \ref{app:wider}.

\subsection{Other two-dimensional systems}
We turn our attention to the guiding effect in other systems. Figures~\ref{fig:fig3}(a)--\ref{fig:fig3}(d) show the conductance as a function of $V_\mathrm{cnt}$ and $V_\mathrm{bg}$ for SLG, 2DEG, BLG, and dtBLG. 
Figures \ref{fig:fig3}(e)--\ref{fig:fig3}(h) in the middle row of the figure present the cross sections of conductance in each system along the red lines for fixed $V_\mathrm{bg}=-18$ V, and the respective band structures are plotted in the bottom row of the panel, in Figs.~\ref{fig:fig3}(i)--\ref{fig:fig3}(l), at selected points marked with the stars in Figs.~\ref{fig:fig3}(e)--\ref{fig:fig3}(h). The band structures are calculated for translationally invariant systems of width 300 nm.
Below we describe the characteristics of each system.

\subsubsection{Semiconductor two-dimensional electron gas}
\begin{figure}[t]
\includegraphics[width=\columnwidth]{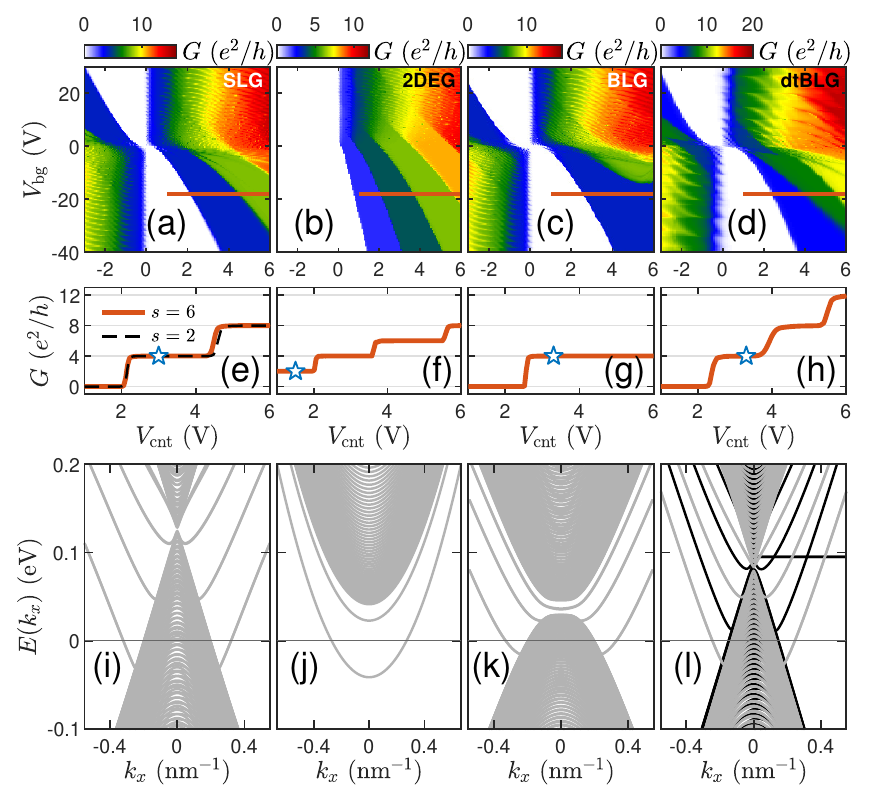} 
\caption{(a)--(d) Two-terminal conductance in SLG, 2DEG, BLG, dtBLG, respectively, as a function of $V_\mathrm{cnt}$ and $V_\mathrm{bg}$, (e)--(h) its cross section at $V_\mathrm{bg}=-18$ V, and (i)--(l) the dispersion relation at a point marked by a star in (e)--(h). The nondispersive band in (l) corresponds to the edge mode in the zigzag-terminated layer of dtBLG. } 
\label{fig:fig3}
\end{figure}

One fundamental difference between the 2DEG and graphene is that for the former, the dispersion relation
does not exhibit the smooth transition between electron- and hole-like conductance. 
For the 2DEG model, the conduction band and the valence band have to be introduced explicitly. 
We focus on transport in the conduction band, $V_\mathrm{cnt}>0$ [see Fig.~\ref{fig:fig3}(b)]. 
The modulation of the potential profile underneath the CNT creates a potential well, leading to electron confinement and formation of discrete modes. The conductance steps at multiples of $2e^2/h$ arise due to the spin degeneracy, in contrast to the quantization at $4e^2M/h $ typical for graphene [see the cross sections in Figs.~\ref{fig:fig3}(e) and \ref{fig:fig3}(f)].
An apparent effect of the confinement seen in the dispersion relation in Fig.~\ref{fig:fig3}(j) is the occurrence of distinct subbands which are well separated from the bulk dispersion relation. The CNT gating allows obtaining well defined conductance steps, offering an alternative to quantum point contacts (QPCs), which can be induced in 2DEG for example by split gates \cite{Wees1988, Lei2021}. 

\subsubsection{Bernal-stacked bilayer graphene}

In Bernal-stacked bilayer graphene, charge carriers are decribed by massive Dirac fermion band structure consisting of two parabolic bands \cite{McCann2013}, with a band gap tunable, e.g., with external gates.
The conductance map in Fig.~\ref{fig:fig3}(c) is obtained with $s_F=4$. It shows quantized steps, 
at $4e^2M/h$ due to the spin and valley degree of freedom, as also seen in the cross section in Fig.~\ref{fig:fig3}(g). 
Conductance quantization in BLG has been previously obtained with external electrostatic gates forming QPCs, however, such an approach requires using a combination of split gates to form a narrow channel and a top gate to tune its density \cite{Kraft2018, Overweg2018}. Our results show an alternative approach by CNT gating, allowing for a simpler device geometry. The lowest plateau with $M=1$
extends over a broad voltage range. This feature can be used to form a quasi-1D BLG chain, robust against the voltage changes, allowing for example investigations of 1D superlattices in BLG. The band structure in Fig.~\ref{fig:fig3}(k) consists of the bulk BLG bands, as well as additional branches due to the CNT-induced channel.

\subsubsection{Decoupled twisted bilayer grahpene}
For dtBLG we consider the top and bottom layer oriented such that the transport direction is along the armchair and zigzag lattice orientation, respectively. This choice corresponds to the relative rotation angle of $30 \degree$ between the sheets, which was found to lead to the interlayer decoupling near the Dirac point \cite{Deng2020, Pezzini2020, piccinini2021parallel}. However, the two graphene sheets are atomically close to each other and the electric charge present on the layers causes effective gating between them \cite{Rickhaus2020electronicthickness, Mrenca2021}. 

The dtBLG device conductance shown in Fig.~\ref{fig:fig3}(d) is calculated for the computational box of size $L\times W=160\ \mathrm{nm}\times 170$ nm, with $s_F=1$. In the bipolar region, $V_\mathrm{bg}V_\mathrm{cnt}<0$, it exhibits two sets of plateaus, dispersing at a different rate with $V_\mathrm{cnt}$ and $V_\mathrm{bg}$. The cross section in Fig.~\ref{fig:fig3}(h) reveals conductance quantized at $4e^2M/h $.
Figure \ref{fig:fig3}(l) shows two overlaid band structures: of the top and bottom layer, plotted with gray and black lines, respectively. The band structure of the bottom, zigzag-terminated layer, is shifted by $k_x=-\frac{2\pi}{3a}$, such that only one of the Dirac cones in the zigzag ribbon band structure is visible in the plot, centered at $k_x=0$. In both band structures one can spot discrete branches detached from the Dirac cones, corresponding to the guiding modes, but due to the electrostatic interlayer coupling described above, the onset of the guiding modes in each layer occurs at a slightly different $V_\mathrm{cnt}$. 
The dtBLG conductance in Figs.~\ref{fig:fig3}(d) and \ref{fig:fig3}(h) is a sum of two individual layers conductance with the steps occurring at different CNT voltage values.

\begin{figure}[t]
\includegraphics[width=\columnwidth]{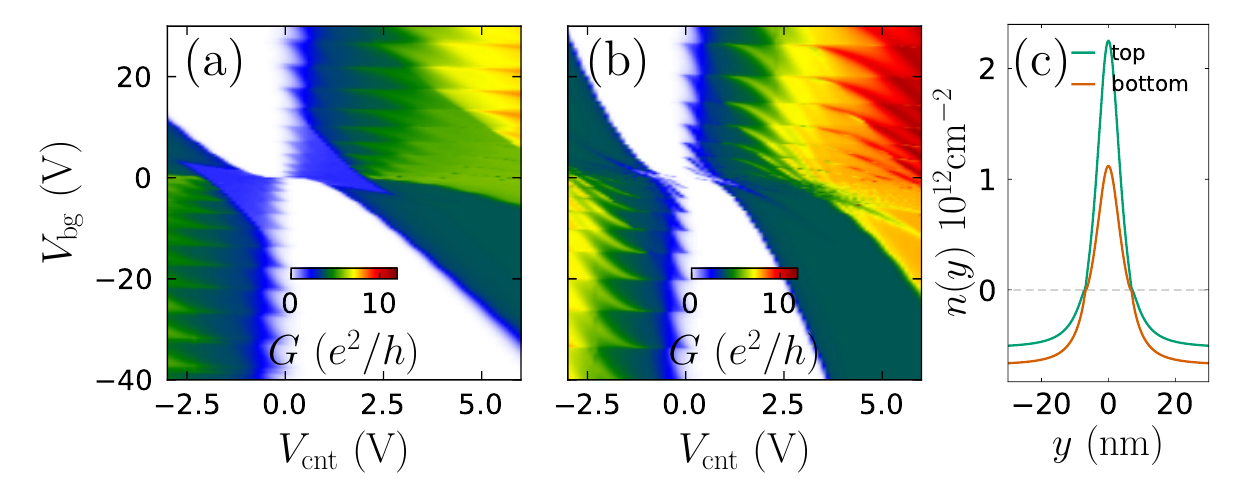} 
\caption{Two-terminal conductance calculated for the (a) bottom and (b) top graphene layer, sum of which yields Fig.\ \ref{fig:fig3}(d). (c) Density profiles of the top and bottom layer at $V_\mathrm{bg}=-18$ V, $V_\mathrm{cnt}=3.3$ V. } 
\label{fig:fig4}
\end{figure}
The individual layers contributions to conductance are shown in Figs.~\ref{fig:fig4}(a) and \ref{fig:fig4}(b). The difference in the dispersion of plateaus in the two layers can be immediately understood when comparing the top and bottom layer density profiles in a representative case of $V_\mathrm{bg}=-18$ V, $V_\mathrm{cnt}=3.3$ V shown in Fig.~\ref{fig:fig4}(c). The carrier density of the top layer is significantly higher than of the bottom one, due to the capacitive coupling between the layers.
In particular, the CNT gates the graphene sample and induces electric charge on the top layer which in turn leads to an effective gating of the bottom layer. This effective CNT gating of the bottom layer is weaker than that of the top layer, leading to different carrier density profiles. To summarize, CNT guiding in dtBLG can be realized in two layers in parallel, such that two independent channels contribute to the conductance, with a full control over the two layers by external gates.

\section{Electron interferometer}

As we have shown in the previous sections, a CNT used as a gate can induce a sharp and narrow waveguide. This, as well as the flexibility of CNTs in terms of their shapes \cite{Wang2006, Joselevich2009}, makes them ideal candidates for building blocks of more complex devices. Here we propose a ringlike Aharonov-Bohm (AB) interferometer and a two-path interferometer. 

When it comes to the characterization of interferometers, their performance is determined by the visibility of the interference pattern, defined as $\alpha=(G_\mathrm{max} - G_\mathrm{min})/(G_\mathrm{max} + G_\mathrm{min})$, where $G_\mathrm{max}$ and $G_\mathrm{min}$ are the maximum and minimum conductance, respectively.
As we show in this section, CNT gating is useful for obtaining conductance oscillation with high visibility in both kinds of interferometers considered here.

\subsection{Aharonov-Bohm interferometer}
\begin{figure}[t]
\includegraphics[width=\columnwidth]{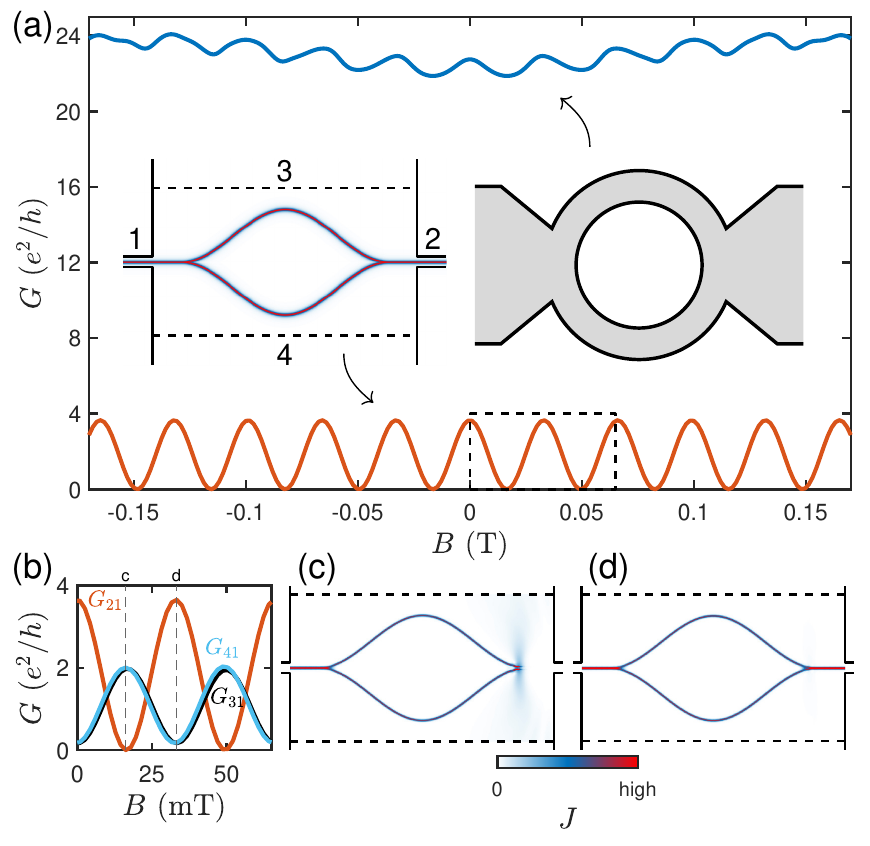} 
\caption{ (a) Two-terminal conductance as a function of the magnetic field $B$ of an etched graphene quantum ring and graphene gated by bent CNTs to form the guided quantum ring, as indicated by the insets showing the ring geometries. (b) The conductance between the injector and the wide top/bottom leads of CNT-gated device in the region indicated by a rectangle in (a). (c)--(d) Current densities at a minimum and maximum of conductance of gated ring, indicated by the vertical dashed lines in (b).}  
\label{fig:fig5}
\end{figure}

We first focus on conductance of an AB interferometer induced by CNT gating and an etched graphene quantum ring. The insets of Fig.\ \ref{fig:fig5}(a) show the geometries of the considered systems. The CNT-gated ring shape is described by a piece-wise function [see Fig.\ \ref{fig:fig5}(a), left inset]
\begin{equation}
y = 
\begin{cases}
0\ , & |x|>\dfrac{W}{2} \\
\pm D\left(\cos\dfrac{2\pi x}{W} + 1\right)\ , & \text{otherwise}
\end{cases}\ ,
\end{equation}
where $W = 500 \sqrt{2}$ nm, $D = 20\pi \sqrt{2}$ nm, and the ring area $A_{c}=2DW=200^2\pi\ \mathrm{nm}^2$. The etched ring [Fig.\ \ref{fig:fig5}(a), right inset] has an inner radius $R_\mathrm{in} = 160$ nm and outer radius $R_\mathrm{out} = 240$ nm, and is attached to leads 400 nm wide. The size was chosen such that the area of a circle of radius $\bar R=(R_\mathrm{in}+R_\mathrm{out})/2=200$ nm, namely $A_r=\pi \bar R^2$, is equal to that of the CNT-gated ring. The choice of the ring geometry refers to a recent experiment \cite{dauber2021}.
The transport calculation was done with scaling factor $s_F=7$ ($s_F=11$) for the CNT-gated (etched) ring. 

\subsubsection{CNT-gated Aharonov-Bohm ring}

For the CNT-gated ring we choose $V_\mathrm{bg}=7.85$ V and $V_\mathrm{cnt}=-2$ V, yielding the densities $n_\mathrm{in} =-4.42 \times 10^{12}\ \mathrm{cm}^{-2}$ and $n_\mathrm{out}=5.3\times 10^{11}\ \mathrm{cm}^{-2}$. This corresponds to $M=1$ [see Fig.\ \ref{fig:fig2}(a)], and the guided current flows from the left narrow lead to the right one, whereas the bulk modes are absorbed by the top and bottom leads.
The CNT-gated ring conductance is shown in Fig.\ \ref{fig:fig5}(a) (orange line) as a function of magnetic field. The oscillation amplitude is nearly $4e^2/h$, with $\alpha=99.87\%$, 
and the oscillation period is $\Delta B\approx 33$ mT which is in agreement with the period evaluated with the area enclosed by the two channels $\Delta B = h/eA_c = 32.9$ mT, confirming that the oscillation is due to the AB effect within the CNT-induced channels. 
At the points of completely destructive interference, the transmission goes mainly to the side drain leads, as evidenced by the high value of $G_{31}$ and $G_{41}$ in Fig.~\ref{fig:fig5}(b). Figures \ref{fig:fig5}(c)--\ref{fig:fig5}(d) show the current densities at selected points indicated by the dashed lines in Fig.~\ref{fig:fig5}(b). The current at the minimums of $G_{21}$ is mainly scattered at the point where the two paths come together close to the right exit, as shown in Fig.~\ref{fig:fig5}(c). By contrast, at the maximum the guided current flows to the right lead without scattering [Fig.~\ref{fig:fig5}(d)].

\subsubsection{Etched Aharonov-Bohm ring}

For the etched ring, we consider carrier density equal to $n= -3.8122\times 10^{12}\ \mathrm{cm}^{-2}$. The conductance as a function of magnetic field is shown in Fig.\ \ref{fig:fig5} (blue line). The high conductance value occurs because of multiple modes propagating in the leads and ring arms. Nevertheless, we can see clear oscillation with a period $\Delta B \approx 33$ mT, same as in the CNT-gated ring, and in agreement with the period expected from $\Delta B = h/eA_r$.

The oscillation visibility $\alpha\approx 1.96\%$ in the etched ring is significantly lower than in the CNT-gated system, and we expect that this can also be the case in experiments. The experimentally measured visibility is expected to be lowered by the interference of multiple paths within the ring arms, as well as by the contact resistance and disorder, in particular the edge roughness introduced in the etching process \cite{Russo2008, Huefner2010, Dauber2017, dauber2021}. 
The current in a CNT-gated channel is confined to a narrow area, thus the carriers pick up nearly equivalent AB phases on their paths in the two ring arms, leading to a perfect destructive interference, as opposed to the etched ring with multiple interfering paths. Furthermore, the edge disorder is excluded in the electrostatically induced ring. Hence, overall the visibility in the experimental CNT-gated ring can exceed that observed in etched rings.

\subsection{Two-path interferometer}

\begin{figure}[t]
\includegraphics[width=\columnwidth]{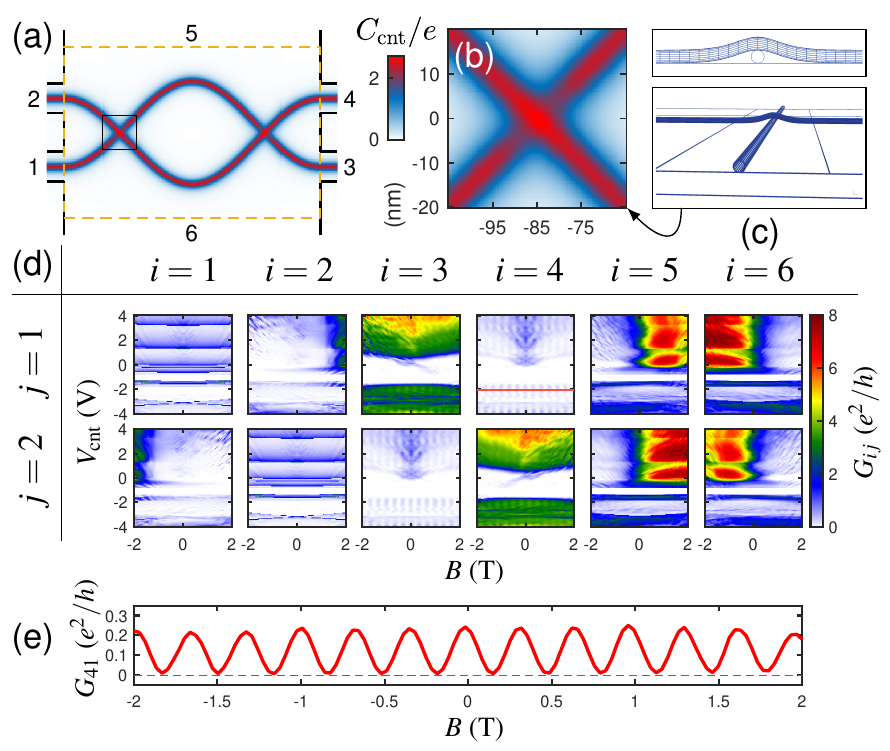} 
\caption{(a) The capacitance profile of the modeled two-path interferometer induced by two crossed CNTs. The solid lines show the region considered for transport calculation, the leads are labeled by numbers 1 to 6, and the dashed lines separate the leads from the scattering region. (b) Zoom of the capacitance profile around the crossing point of the CNTs within the square marked in (a). (c) Geometry of the crossed CNTs adopted for the finite element method electrostatic simulation. (d) Conductance between pairs of leads labeled in (a). (e) Cross section of $G_{41}$ in (d) along the red line at $V_\mathrm{bg}\approx-2$ V.}
\label{fig:fig6}
\end{figure}

Interferometers proposed recently in graphene rely on beam splitting
at the pn junction close to the lattice termination \cite{Tworzydlo2007, Morikawa2015, Mrenca2016, Wei2017, Handschin2017, Jo2021} or with the aid of the insulating $\nu=0$ quantum Hall state \cite{Deprez2021, Ronen2021}, and require high magnetic fields such that the quantum Hall edge and pn junction states are formed.
In this section, we  propose a two-path interferometer with a fully electrostatic beam splitter at the crossing of the two guiding channels,
induced by two bent CNTs placed on top of each other. Figure \ref{fig:fig6}(a) shows the considered geometry. 
The electron beam from one of the injectors is split at the crossing between the two channels. The two paths traversed by the electrons encircle a closed area, and at the other crossing either beam can end up in one of the detector leads on the right. The magnetic flux through the area enclosed by the two paths results in a phase difference between them which gives rise to the conductance oscillation.

The two paths are described by the functions $\pm [D \cos(2 \pi x / W) + h]$, with $D=50$ nm, $W=300$ nm, and $h=10$ nm. For modeling  of a realistic device we take into account that at the crossing point, within $x\in (-105, -65)$ nm and $x\in (65,105)$ nm [see the area marked by a square in Fig.\ \ref{fig:fig6}(a), and enlarged in Fig.\ \ref{fig:fig6}(b)] one CNT lies on top of the other and bends locally [see Fig.\ \ref{fig:fig6}(c)] \cite{Avouris1999}. The capacitance within the square area is obtained from a 3D finite-element electrostatic simulation, and the resulting $C_\mathrm{cnt}(x,y)$ [shown in Fig.\ \ref{fig:fig6}(b)] is then combined with the straight CNT capacitance. The  overall capacitance profile is shown in Fig.\ \ref{fig:fig6}(a). 

Figure \ref{fig:fig6}(d) shows the conductance $G_{ij}$ from leads $j=1,2$ to $i=1, \dots, 6$ [as labeled in Fig.\ \ref{fig:fig6}(a)], as a function of magnetic field and $V_\mathrm{cnt}$, for a fixed $V_\mathrm{bg}=7.85$ V. We notice that the conductance between bottom left to bottom right ($G_{31}$) and upper left to upper right ($G_{42}$) is high since  in this case the current guided by the channel follows a smooth trajectory and goes preferably along the straight part at the crossing point of the CNTs. Splitting of the current at the crossing still occurs and gives rise to oscillating conductance $G_{41}$ and $G_{32}$ with an amplitude $\approx$ 0.2 $e^2/h$ [see the line cut in Fig.\ \ref{fig:fig6}(e) corresponding to the red line marked on the $G_{41}$ map of Fig.\ \ref{fig:fig6}(d)]. The oscillation is due to the magnetic flux piercing the area enclosed by the two crossing channels, and the oscillation period $\Delta B\approx 0.33$ T is in agreement with the one evaluated given the loop area, $\Delta B = h/eA=0.324$ T. The visibility of the $G_{41}$ and $G_{32}$ oscillation $\alpha=95.04\%$ is high although the relatively low amplitude corroborates that the beam splitting is asymmetric. This can be improved by decreasing the crossing angle between the channels.
A strong asymmetry of the conductance $G_{51}, G_{61}, G_{52}, G_{62}$ at $V_\mathrm{bg}\gtrsim 0$ [i.e., unipolar regime -- cf. Fig.\ \ref{fig:fig2}(a)] occurs because the bulk modes injected from the left leads preferably flow to the lead 5 (6) for positive (negative) magnetic field due to the Lorentz force. 
The results shown here suggest the crossed CNTs can be used to work as electrostatic beam splitters that operate in moderately weak to strong magnetic fields.

\section{Diffractive point injector}

\begin{figure}[t]
\includegraphics[width=\columnwidth]{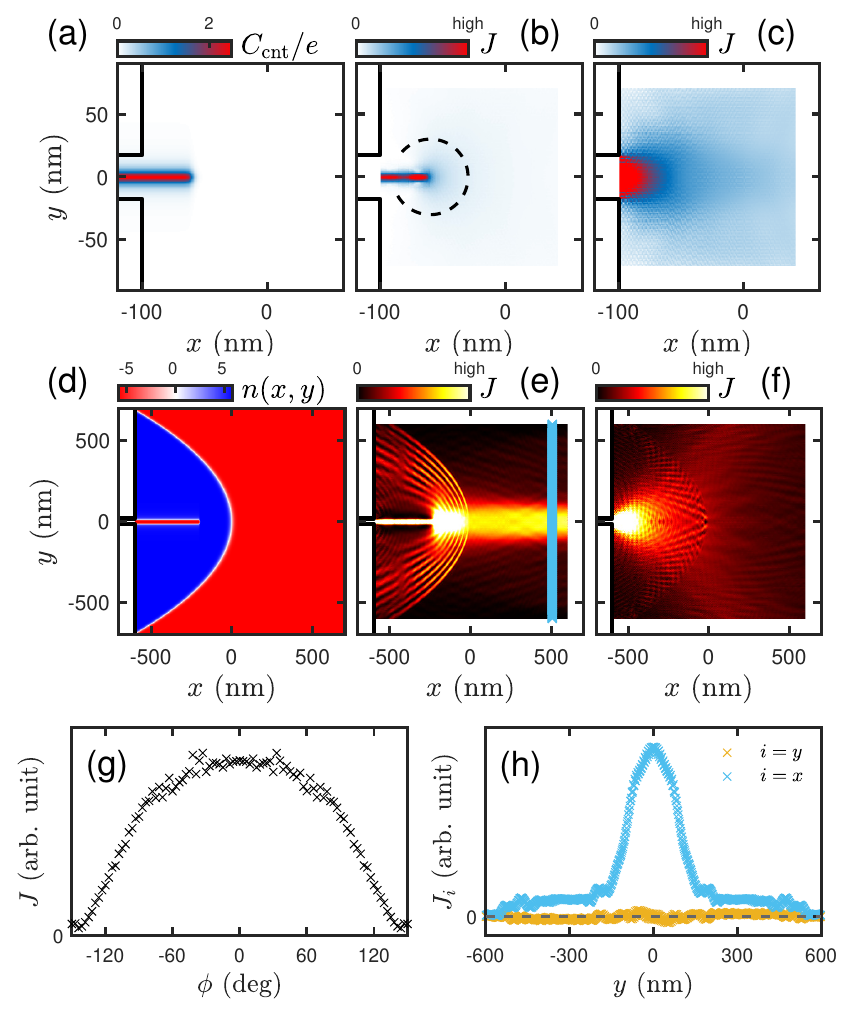} 
\caption{(a) Modeled capacitance between the truncated CNT gate and graphene. (b),(c) Current density maps at $V_\mathrm{bg}=7.39$ V and $V_\mathrm{cnt}= -2$ V, with and without the CNT gate, respectively. (d) Carrier density of the lensing apparatus composed of a CNT gate truncated at the focal point of a parabolic interface where the symmetric pn junction forms. (e),(f) Current density maps with and without a CNT gate, respectively.
(h) The vector components of current along the blue line cut in (e). (g) The angular distribution of current along the dashed line in (b) where radius $r = 30$ nm, with respect to the terminal point. } 
\label{fig:fig7}
\end{figure}

\subsection{Truncated CNT gate}
Point contacts in graphene are becoming an essential component in a number of electron optical application such as Dirac fermionic optics cavities \cite{Schrepfer2021dirac} and electron collimation \cite{Liu2017beams, PhysRevB.100.045137}. In particular,  approaching the limit of quantum-to-classical correspondence of the focused electron
waves \cite{Liu2017beams} requires a pointlike injector. However, the state-of-the-art experimental realization of point contacts is limited to 100 nm in diameter currently, using the prepatterning of the top hBN \cite{doi:10.1063/1.4935032}. We propose an alternative scheme for a pointlike injector using a truncated CNT gate. The termination of the highly spatial-confined channel in graphene induced by the truncated CNT gate naturally forms 
an electron point injector that scales down to the order of 10 nm. For the modeling of the electrostatic coupling between the truncated CNT gate and graphene, we use $s_F=7$, and consider the capacitance in Fig.\ \ref{fig:fig1}(c), multiplied by a smoothness function, $(1+\tanh((x_\mathrm{truncation}-x)/d_\mathrm{smooth}))/2$, where $x_\mathrm{truncation} = -60$ nm and $d_\mathrm{smooth} = 3$ nm, as shown in Fig.\ \ref{fig:fig7}(a). An example of the current density for $V_\mathrm{bg}=7.39$ V and $V_\mathrm{cnt}= -2$ V, where a single mode emerges in the guiding channel, is shown in Fig.\ \ref{fig:fig7}(b). The small section of the lower current density in the channel results from the standing wave due to the partial reflection from the truncation point. The angular distribution of the current density in Fig.\ \ref{fig:fig7}(g) shows its directional characteristic: higher intensity is seen at small angles, as opposed to an ideal point injector which does not show any directional dependence. Also, in contrast to the current injected from a lead [Fig.\ \ref{fig:fig7}(c)], the truncated CNT forms a current source of a smaller size. Nevertheless, for certain applications a uniform current distribution is not crucial, one of the examples being collimation of electron beams, described in the next subsection. 

\subsection{Generating electron beam}
To further demonstrate the utility of the truncated CNT gate, 
we combine the pointlike injector with a parabolic-shaped pn junction to form an electron beam generator, following Ref.\ \onlinecite{Liu2017beams}. The difference here is that the role of the pointlike injector is played by the CNT gate truncated at the focal point of the parabolic pn junction, instead of a pointlike contact \cite{Liu2017beams, Klein1929}. The pn junction symmetric in the carrier density is modeled by a smooth function, $n_\mathrm{junction} \tanh[(x_\mathrm{parabola}-x)/d_\mathrm{smooth}]$ describing the $x$ dependence and a quadratic function and $x_\mathrm{parabola} = -y^{2}/4f$ accounting for the $y$ dependence, where the carrier density $n_\mathrm{junction} = 5.78 \times 10^{11}\ \mathrm{cm}^{-2}$,  the smoothness parameter $d_\mathrm{smooth} = 15$ nm, and the focal length $ f = 200$ nm. 

Figure\ \ref{fig:fig7}(d) shows an exemplary carrier density $n(x,y)$ considering $V_\mathrm{cnt} = -3.5\unit{V}$. The resulting current density map is shown in Fig.\ \ref{fig:fig7}(e), where a well collimated electron beam at the right side of the parabolic pn junction can be seen. 
The generated electron beam, as explained already in Ref.\ \onlinecite{Liu2017beams}, is a consequence of the negative refraction combined with the Klein collimation \cite{Cheianov2006}, which describes the transmission function that decays with the incidence angle, from perfect at normal incidence, known as Klein tunneling \cite{Cheianov2006, Katsnelson2006, Klein1929}, to zero at a certain finite angle depending on the smoothness of the pn junction \cite{Cheianov2006}.
The nearly perfectly collimated electron beam is further examined by showing the $x$ and $y$ components of the 2D current density $\mathbf{J} = (J_x,J_y)$. Figure\ \ref{fig:fig7}(h) shows the line cut of $J_x$ and $J_y$ along the path marked on Fig.\ \ref{fig:fig7}(e). The vanishing $J_y$ shows the efficient collimation of the current as a consequence of negative refraction of the pointlike source positioned at the focal point of the parabolic pn junction. The collimated current density at the right side of the junction exhibits a $J_x$ distribution that peaks around the parabola axis at $y=0$, as a consequence of the Klein collimation.  
On the other hand, current injected directly from a lead without a CNT is not collimated, as shown in Fig.\ \ref{fig:fig7}(f).

\section{Concluding remarks}
In summary, we investigated the guiding effect in CNT-gated two-dimensional systems, and found well-defined conductance plateaus when discrete guiding modes and bipolar junctions -- in graphene-based systems -- are formed in the channel. This mechanism of conductance quantization in SLG is an alternative to QPCs which were so far created by etching graphene rather than gating it due to the difficulty in inducing a bandgap in graphene. The conductance plateaus obtained by CNT gating are sharper compared to the plateaus in etched graphene QPCs as well as other systems, which can be electrostatically depleted to form QPCs, including BLG and 2DEG. Moreover, CNT guiding works well in curved channels, making them useful as building blocks for electro-optical devices, including quantum rings and other interferometers. Thanks to the character of carrier confinement in CNT gated channels, they are not limited to the operation at strong magnetic fields, as opposed to interferometers based on pn junctions or the insulating $\nu=0$ quantum Hall state. As presented here, this can also be used to create point injectors simply by gating with a CNT with an abrupt termination. 
CNT gating allows electrostatic confinement of carriers which is a way to exclude imperfections like edge roughness introduced in the etching process in lithographically defined gates, offering a versatile tool for electro-optical components.

\begin{acknowledgments}
Financial support from Taiwan Ministry of Science and
Technology (109-2112-M-006-020-MY3) is gratefully acknowledged. This
research was supported in part by PL-Grid Infrastructure and Higher Education Sprout Project, Ministry of Education to the Headquarters of University of Advancement at National Cheng Kung University.
\end{acknowledgments}

\appendix

\section{Curved CNT capacitance profile}
\label{app:C1Dto2D}
\begin{figure}[tbh]
\includegraphics[width=0.6\columnwidth]{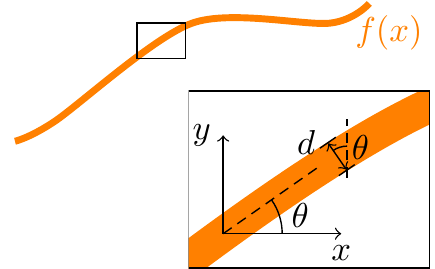} 
\caption{Sketch of the approximated capacitance induced by a curved CNT with the shape described by a function $f(x)$. } 
\label{fig:figSA}
\end{figure}

In Sec.\ III we consider devices gated with curved CNTs. For simplicity and to avoid the need for electrostatic simulation for each system, the capacitance of a
 curved CNT $\tilde C_\mathrm{cnt}(x, y)$ is calculated using $C(y)$, the transverse capacitance profile of a straight CNT, as illustrated in Fig.~\ref{fig:figSA}. 
For a CNT shape described by a function $f(x)$, the original 1D profile is shifted in the $y$ coordinate, and next, to take into account the 
channel slope $f'(x)=\tan \theta$, it is scaled by $\cos\theta$. The resulting formula reads
\begin{equation}
\tilde C_\mathrm{cnt}(x, y)=C_\mathrm{cnt}\textbf{(}\cos \theta(x)[y - f(x)]\textbf{)}.
\end{equation}

\section{Disorder}
\label{app:disorder}

\begin{figure}[tbh]
\includegraphics[width=\columnwidth]{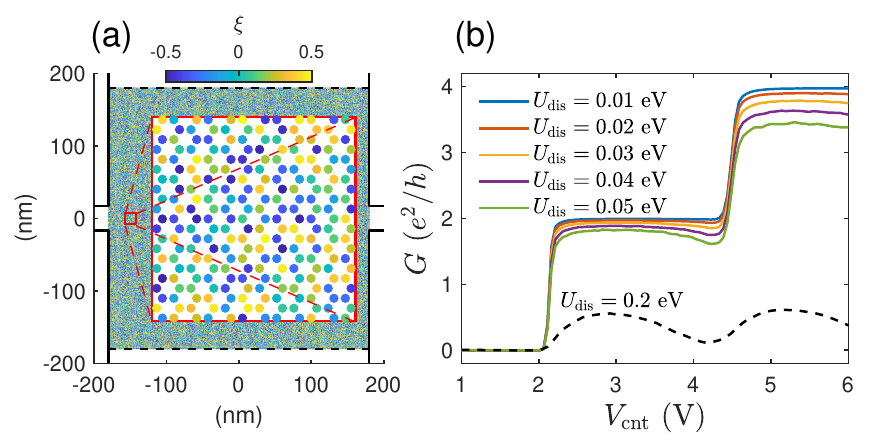} 
\caption{(a) One of the random configurations of the disorder parameter $\xi$. (b) Conductance line cut at $V_{\mathrm{bg}}=-18$ V and $s_F=6$ with a few disorder strength values $U_{\mathrm{dis}}$. } 
\label{fig:figSB}
\end{figure}

To check if the guiding effect is robust against disorder, we 
added an on-site potential $\xi U_{\mathrm{dis}}$, with $\xi$ being a random number, $\xi \in (-0.5, 0.5)$ [see Fig.\ \ref{fig:figSB}(a)], and $U_{\mathrm{dis}}$ is the maximum disorder strength. We calculated the conductance averaged over 200 disorder configurations for a fixed $U_{\mathrm{dis}}$ (for $U_{\mathrm{dis}}=0.2$ eV, we used 1000 configurations). Figure \ref{fig:figSB}(b) shows the conductance for $V_{\mathrm{bg}}=-18$ V and $s_F=6$ [see Fig.\ \ref{fig:fig3}(e) in the main text]. For moderate disorder with $U_{\mathrm{dis}}\lesssim 0.05$ eV the plateaus are slightly disturbed but close to the expected value $G=4e^2M/h$. However, for considerably strong disorder of the order of $U_{\mathrm{dis}}=0.2$ eV, the steps are destroyed. Disorder of few meV has been observed in graphene samples encapsulated in hBN \cite{Xue2011}, so the present results allow us to conclude that the quantization can be observed in realistic samples. 

\section{Intervalley scattering}
\label{app:plato}
\begin{figure}[tbh]
\includegraphics[width=0.8\columnwidth]{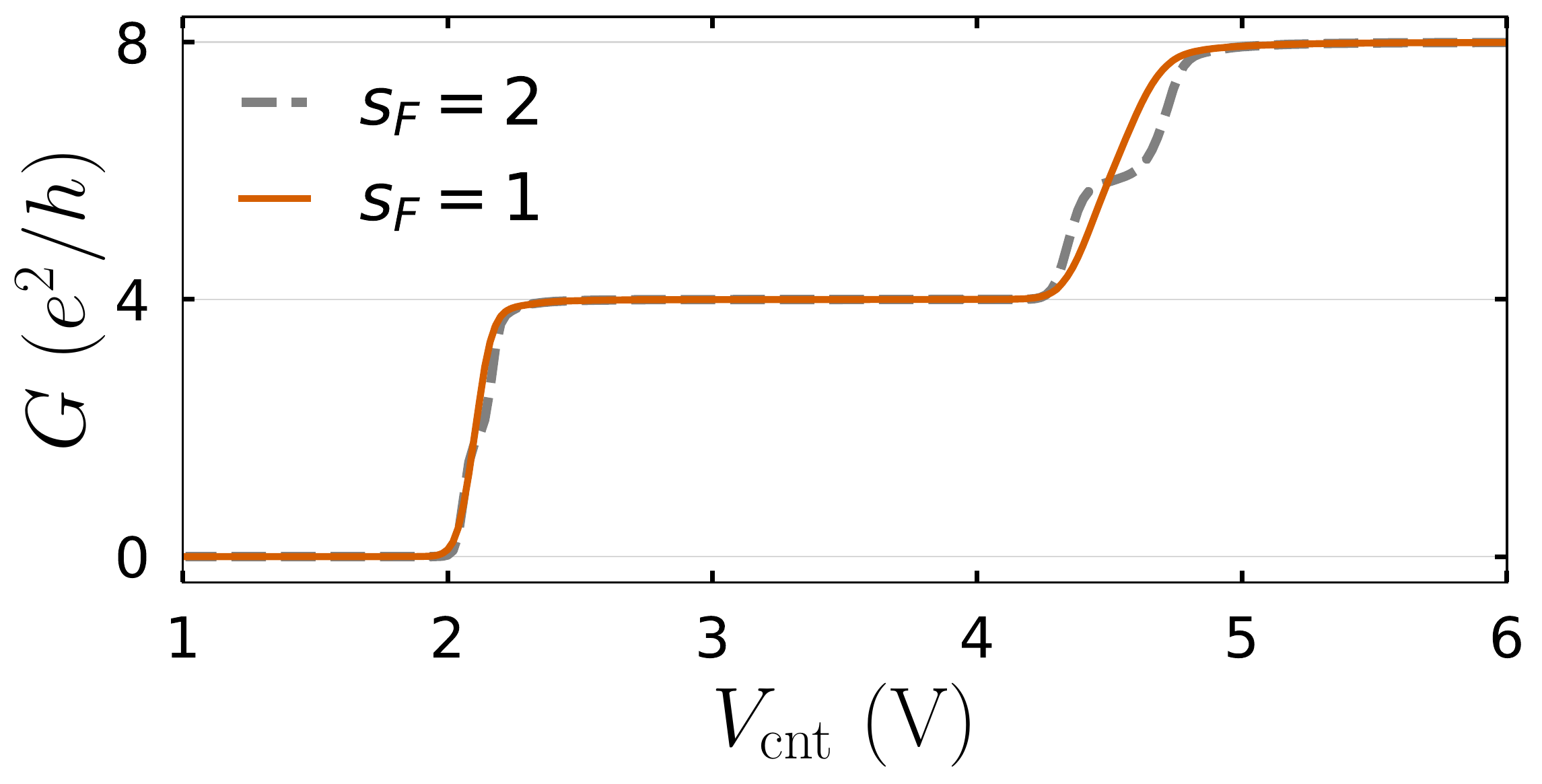} 
\caption{Conductance line cuts with the channel along the zigzag direction for $V_{\mathrm{bg}}=-18$ V with $s_F=1$ and $s_F=2$. } 
\label{fig:figSC}
\end{figure}

For a potential that varies strongly on the length scale of a lattice spacing, the intervalley scattering is present. The CNT-induced potential profile is not sharp enough to cause the intervalley scattering for realistic gate voltages. However, for high gate voltages, in the case of scaled lattice, the potential variation over the lattice spacing is strong. As a result, in a channel induced along the zigzag direction we observe intermediate conductance plateaus at $G=4e^2(M-1/2)/h$, $M=1,2,\dots$, as seen in a conductance line cut for $V_{\mathrm{bg}}=-18$ V presented in Fig.~\ref{fig:figSC}, for $s_F=1$ and $s_F=2$. 
 
\section{Thicker hBN}
\label{app:thick}
\begin{figure}[tbh]
\includegraphics[width=\columnwidth]{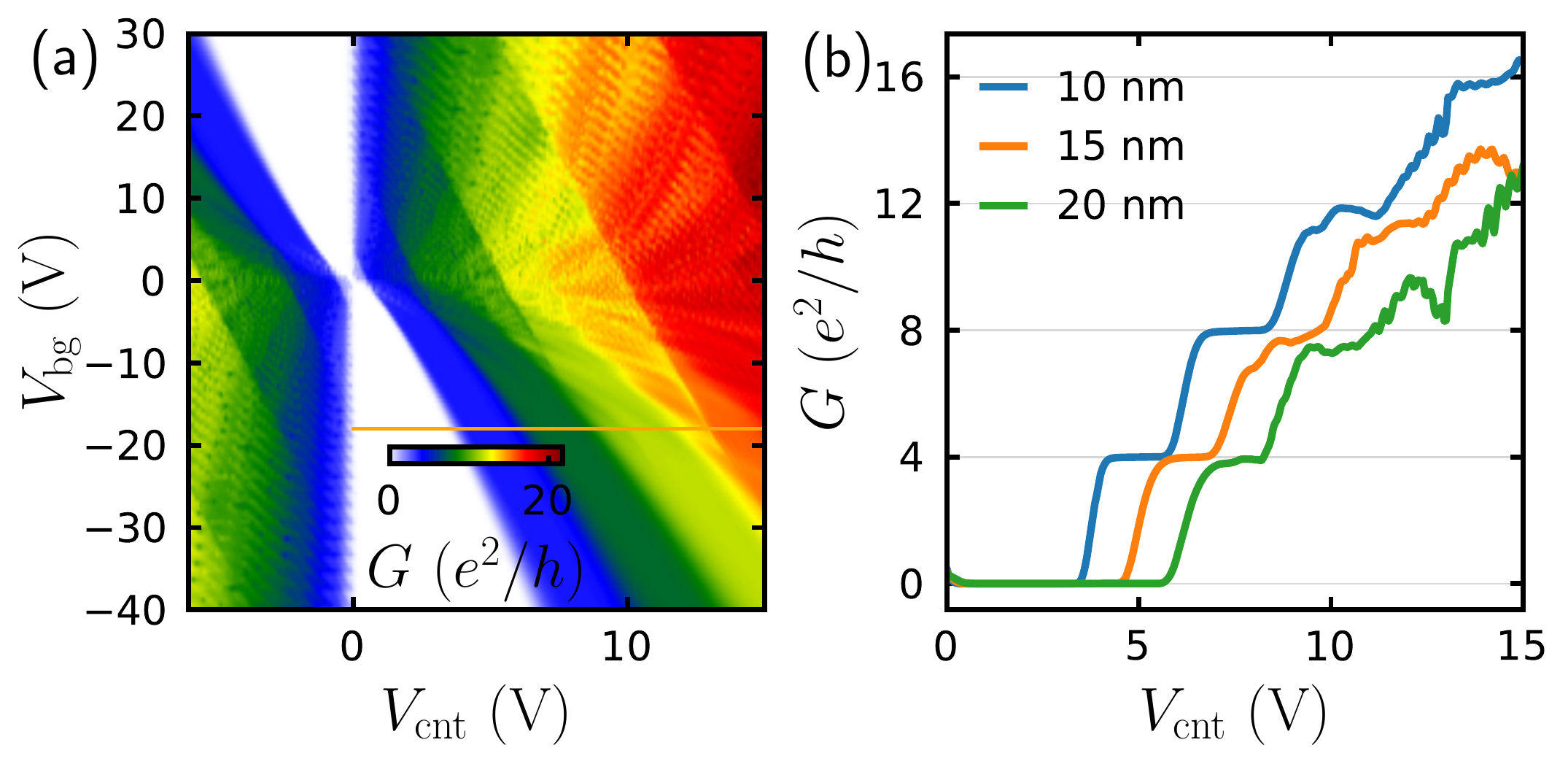} 
\caption{(a) Conductance between thin leads of SLG
as a function of $V_{\mathrm{cnt}}$ and $V_{\mathrm{bg}}$ 
 with 10 nm thick hBN between the CNT and graphene. (b) Line cuts with $V_{\mathrm{bg}}=-18$ V for systems with hBN thickness 10, 15 and 20 nm. } 
\label{fig:figSD}
\end{figure}

For a direct connection with the experimental work in Ref.~\onlinecite{Cheng2019cntguiding}, we consider 4 nm hBN layer between the CNT and graphene. However, for a safer design, a thicker hBN is required to prevent  the possible dielectric breakdown.
We performed calculations for the hBN thicknesses of 10, 15 and 20 nm, for which the breakdown field was shown to be about 13 MVcm$^{-1}$, 12 MVcm$^{-1}$, and 11 MVcm$^{-1}$, respectively \cite{Hattori2015}. The breakdown voltage corresponds to about 14 V, 19 V, and 22 V, respectively, and a wide gate voltage range considered in this work is more reasonable experimentally.
The conductance map obtained for the 10 nm thick hBN is presented in Fig.~\ref{fig:figSD}(a). The conductance plateaus are present up to about $M=3$ for moderately low gate voltages. The line cuts for $V_{\mathrm{bg}}=-18$ V for the three cases are presented in Fig.~\ref{fig:figSD}(b). The conductance quantization is visible, although the steps become smoothened for thicker hBN.

\section{Wider injection lead}
\label{app:wider}
\begin{figure}[t]
\includegraphics[width=\columnwidth]{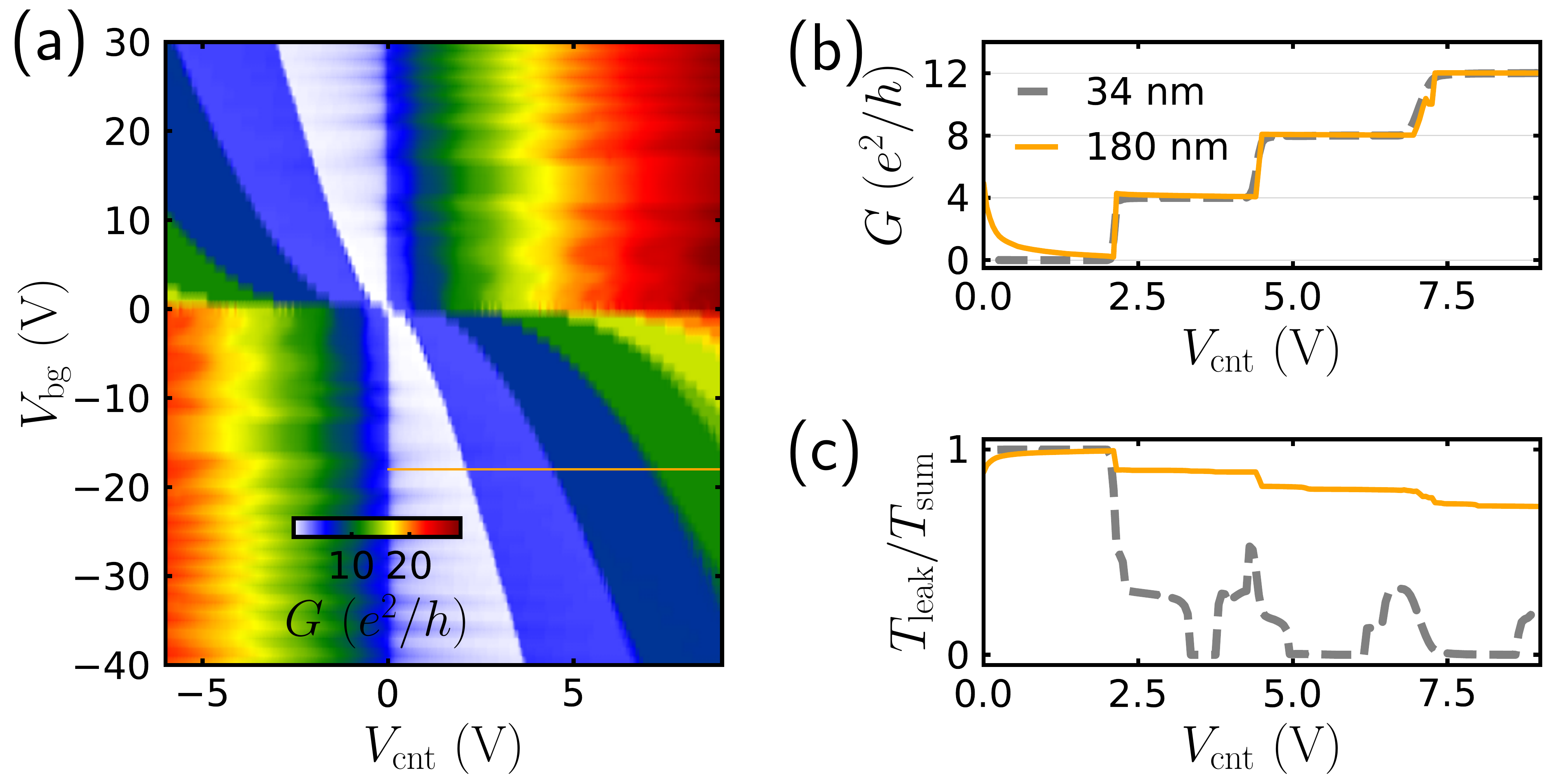} 
\caption{(a) Conductance as a function of $V_{\mathrm{cnt}}$ and $V_{\mathrm{bg}}$ 
 with 180 nm wide injection lead. (b) Line cut with $V_{\mathrm{bg}}=-18$ V. } 
\label{fig:figSE}
\end{figure}

Throughout this work we used narrow injector and collector leads. Here, we consider a system with the lead width $w=180$ nm, and the computational box size $L\times W=1000\ \mathrm{nm}\times 360$ nm. In the conductance map 
 in Fig.\ \ref{fig:figSE}(a) the plateaus are present despite the large lead width. This is better seen in Fig.\ \ref{fig:figSE}(b) which shows the conductance line cuts at $V_{\mathrm{bg}}=-18$ V for $w=180$ nm and $w=34$ nm for comparison. Figure \ref{fig:figSE}(c) presents the ratio of the transmission to the top and bottom drain leads $T_{\mathrm{leak}}$ to the transmission summed over the drains and right lead $T_{\mathrm{sum}}$. Whereas in the narrow lead the leakage may nearly drop to zero, in the wide lead the leakage is high because of the large number of bulk modes. However, the bulk modes could carry additional currents to the right lead, if the leads are very wide and the system is short, such that the conductance plateaus are no longer intact. 

\bibliography{guiding}

\end{document}